# MOBISPA: A REFERENCE FRAMEWORK FOR MOBILE AS A SECURE PERSONAL ASSISTANT

Rakhi Misuriya Gupta
Distinguished Architect – Open Group,
Technology Manager – Tata Consultancy Services.
New Delhi
rakhi.gupta@tcs.com

## ABSTRACT

Mobile is taking center stage and becoming the device of preference for all aspects of communication because of our increasingly "on the go" lifestyles. With this the demands on mobile's capability to execute increasingly complex operations are also on the rise. However, despite improvements in device computing power in the last couple of years a mobile device continues to have limitations. Mobile driven everyday use cases are increasingly raising expectations that rest on mobile technologies that are still evolving. A number of fragmented approaches/solutions have been created that address various requirements unique to mobility, however there is a lack of a single framework that serves as a unifying reference for industry and solution architectures. The paper addresses this concern through the specification of a comprehensive reference framework for mobility that is generic and vendor neutral.

**Keywords**—Mobile Cloud Computing, Mobile Security, Application Delivery Controller (ADC), Network Function Virtualization (NFV), Software Defined Networking (SDN), Composite Capabilities/Preference Profiles (CC/PP), User Agent Profile (UAProf), Mobile Experience Management (MEM), CloneCloud, Cloudlets, Predictive Provisioning, Mobile backend as a Service (mBaaS)

## 1. INTRODUCTION

In the last decade since the arrival of smart phones and the increase in the mobile customer base the mobile revolution has gained momentum. Retail/eCommerce businesses substantial revenue is now increasingly driven by mobile apps. These are not the only cases transforming the mobility landscape, an increasing demand is being placed on mobile service from social apps front as well as from banking apps.

The mobile computing landscape has evolved dramatically, driven at a fast pace through the demands placed by the user and businesses. To differentiate their products vendors are incorporating evolving technologies innovations based on Application Device Controller, Mobile Cloud Computing, Network Function Virtualization and Software Defined Function etc. into their platforms, while research on these areas are still underway. Thus, while comprehensive standards and frameworks are yet to be defined a proliferation of one-of-solutions is already happening that will lead to incompatibility and lack of reuse.

We define an overall reference framework for Mobile Computing that serves to address stakeholders concerns, standards compliance concerns, use case concerns, enterprise architect concerns and infrastructure architecture concerns. The defined framework "Mobile as a Secure Personal Assistant – MOBISPA" is a framework that is generic vendor neutral framework and is intended to act as a reference for design of mobility solution/product, assessment of existing mobility solution or mobility architecture and in the section of a mobility platform.

We also examine the various industry trends and academia research in the context of this framework and incorporate mature standards as building blocks in addition we also identify areas where standards are still developing or need to be explored further.

# 2. DETAILS OF PAPER

A. The Mobile Maturity Curve

There are challenges that are very unique to Mobile computing:

- Limited on device resources in terms of data storage, computation, battery, bandwidth etc.)
- Limited and/or low bandwidth in case of wireless networks resources and even more constrained than wired networks
- Intermittent service due to network congestions, network failures, mobile signal strength problems, dropped connections
- Highly heterogeneous devices – leading to apps portability, interoperability and usability issues. Other issues with mobile device applications
- Highly heterogeneous networks
- Applications for other devices require adaptation both in terms of content and user interface due to the small Screen with low resolution
- Security issues are more profound due to the distributed nature of the network and the device itself
- Multi-tasking and information exchange or seamless mobile computing experience non-existent, with each application being a silo

Despite the challenges mobility has experienced more rapid changes in the last decade than in the entire history of mankind, driven by rapid adoption. Starting 1999, only 15 percent of the world population had telephone to 2009 where more than 70% had mobile phones. Other trends like socialization and retail eCommerce have been instrumental in the growth of OTT (Over-The-Top) mobile services.

No matter where we work in the office or at home in the new millennium, "always connected" is now a key requirement. E-mails are insufficient and productivity depends upon anytime and anywhere access. Across the enterprise processes are being transformed by a move toward mobile workspaces, consumerism is driven by the same mobile device for personal and for business purposes.

To achieve future state functionality a lot more capabilities in the underlying technologies and standards need to evolve rapidly. We map the evolution of the mobile in Figure 1 from the stage of "Device Excitement" in 2008 to today's standard where "Interactive Devices" and "Mobile Experience Management" has become the norm, to the future where mobile will be the ubiquitous personal assistant.

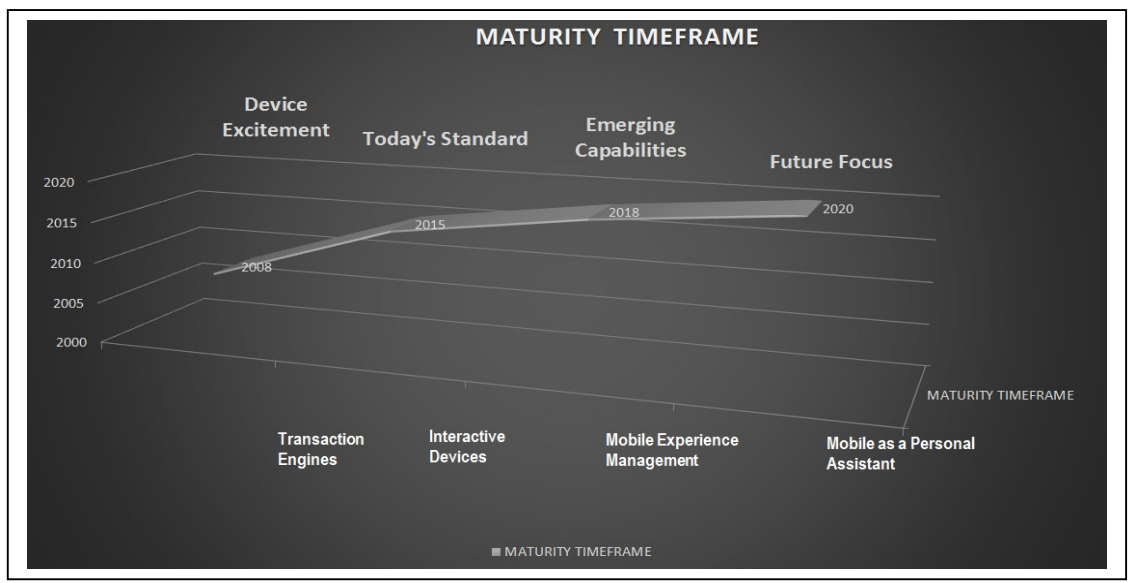

Fig. 1. The Mobile Maturity Curve

## B. A Reference Framework for Mobile as a Secure Personal Assistant

We define a reference framework with elements that address the following key concerns:

- Stakeholder concerns - such as implementers, product vendors, integrators
- Standards compliance concerns- Reference against which standards bodies can extend existing standards, or integrated existing ones)
- Use Case Concerns - Acts as a use cases for Design of mobility solution, assessment of an existing mobility solution or architecture, Mobility Product and Mobility Platform selection.
- Enterprise Architecture Concern- Provides a uniform way of designing solutions to introduce consistency, uniformity and repeatability in the structure of the mobility solutions across the enterprise driven through the need for predictability. It also defines the governance polices and the skill sets needed.
- Infrastructure Architecture Concerns- Key concerns and capabilities required for mobility enablement

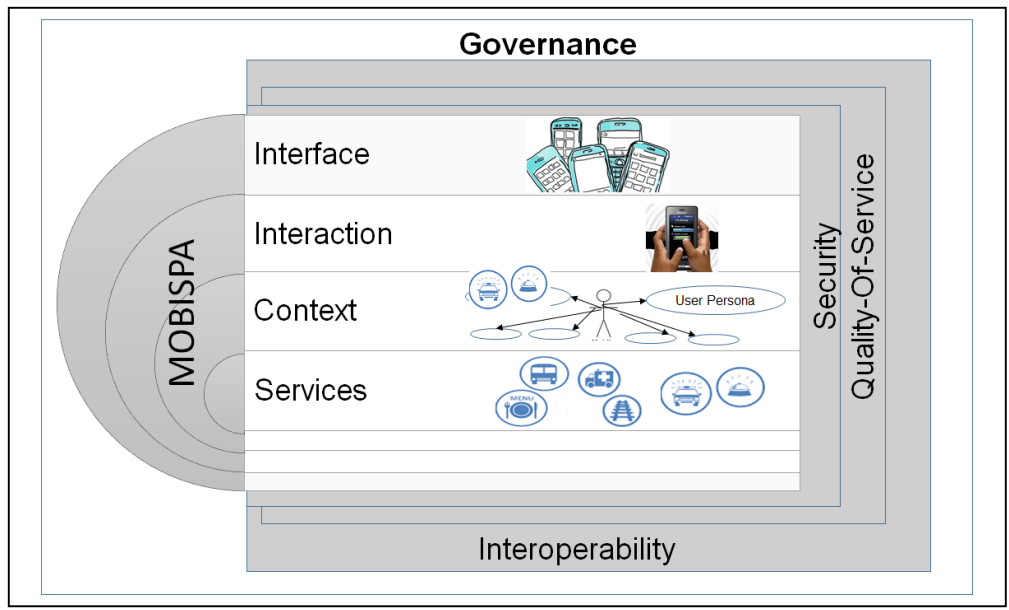

Fig. 2. Reference Framework for Mobile as a Secure Personal Assistant (MOBISPA)

The key areas of concern as identified in Figure 3 above are mapped to various capabilities required in the below table.

Table. 1. Mapping areas of core concern to capabilities desired

| Core Areas of Concern | Capabilities Desired |
|---|---|
| Interfaces | - Application and Information Templates - Visual Styles and aesthetics, Device Page Organization/Layout, Navigation, Display<br>- Application and Information View and Model - Web App / Browser, Complex Data Organization, Social Media Organization, Advanced interaction interface – gesture-prompts, glance-alerts<br>- Device Agnostic Information and Application Structure Content Organization – Abstract User Interface<br>- Accessibility Tools for Display Rendering - Device Native Format Concrete User Interface |

| Core Areas of Concern | Capabilities Desired |
|---|---|
| Interaction | - Multi-Modal Recognition
- Semantic Interpretation
- Interaction Manager |
| Context | - Context Derivation Models - Multi-Dimensional User Persona, Predictive, Autonomous Machine Learning
- Context based Inference
- Context based event generation |
| Mobility Services | - Mobile Backend as a Service enabled through various API's such as – Mobile API's, Enterprise System API's, Data Access API's, Infrastructure API's, LOB API's. As well as other capabilities such as: Push notifications, workflows, ability to automatically generate interfaces (REST based) etc
- Services collaboration – alert, subscription, interoperation, push notifications
- User management
- Data management |

Additionally, certain supporting areas of concern are mapped to capabilities in the below table.

Table. 2. Mapping areas of supporting concerns to capabilities desired

| Areas of Concern | Capabilities Desired |
|---|---|
| Quality-Of-Service (QoS) | Availability, Reliability and Response Time of Computation and Connectivity determine the QoS on the mobile. |
| Security | - Device Security- MDM, Password Policy, Remote wipe
- Application Deployment Security – app/enterprise, store, OTA, provisioning
- Application Data Security – On device data encryption
- Transmission Security – Secure http, x.509 certificates
- Enterprise Security – SSO, LDAP
- Network Security – Middleware Deployment, DMZ, VPN, Reverse Proxy |
| Interoperability | - Any App/Browser any Device
- App as a Service – Adherence to "Mobile backend as a Service" (mBaaS) and therefore also enable app to app communication |

The components that address the various capabilities covered above maybe distributed or centralized in nature depending on the availability of the resources.

The areas of concern span across the device, the distributed nodes on the network, the network itself, hosted infrastructure for backend processing etc. End-to-End Governance is a key aspect that requires extensive research, therefore we omit discussions in the scope of this paper, where we elaborate on the core capabilities and their building blocks in detail and delve on the supporting capabilities to a limited extent to explain the trends and evolution in the area.

C. **Mobile Interface Capabilities**

The mobile interface capability can be represented by the key building blocks

- Application and Information Templates
- Application and Information View and Model
- Abstract User Interface
- Device Native Format Concrete User Interface

Figure 3 below depicts the building blocks.

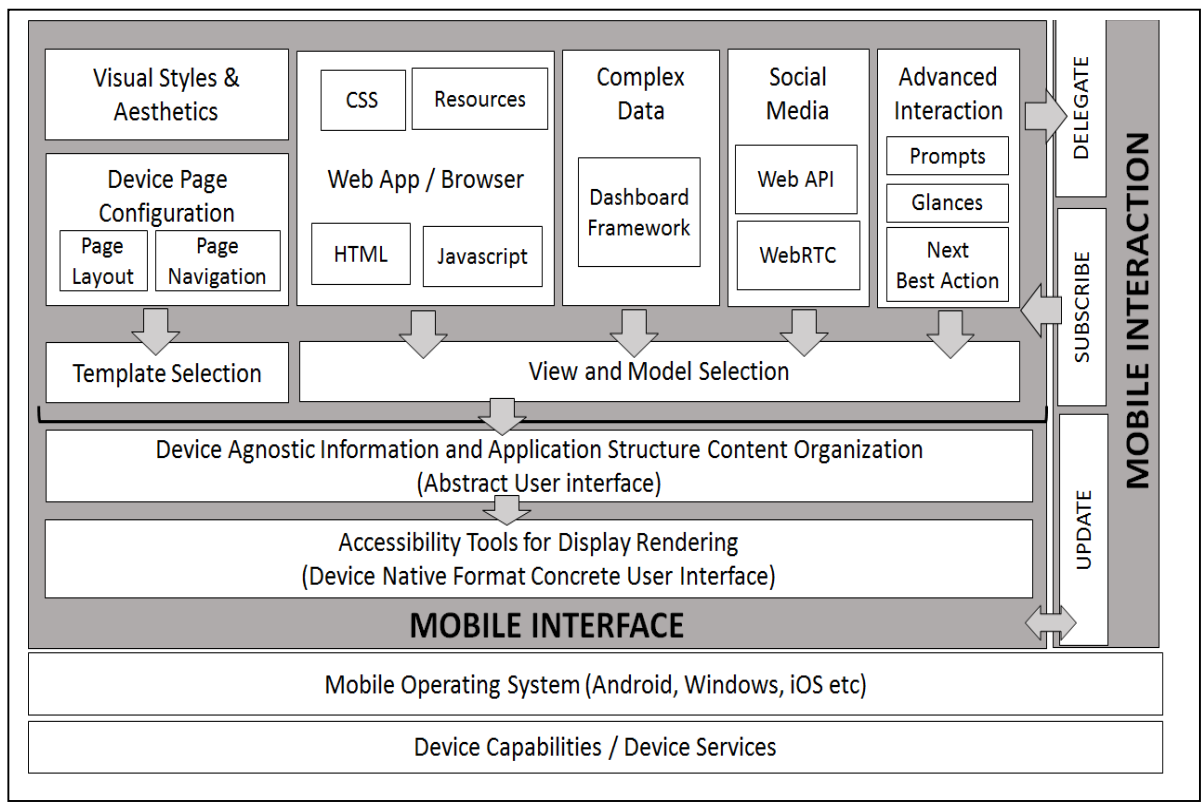

Fig. 3. Building blocks for Mobile Interface Capabilities

The above diagram depicts the flow between the "Mobile Interface" and "Mobile Interaction" capability layers. For more details related to WebRTC reference [23].

### D. Mobile Interaction Capabilities

The building blocks as defined in the multi-modal mobile interaction architecture as defined by W3C can be used as a reference and consists of:

- Multi-Modal Recognition
- Multi-Modal Semantic Interpretation
- Multi-Modal Integration with the Interface via, an interaction manager
- Interaction Manager communicates asynchronously via subscription / delegation services with Mobility Interface and synchronously via back and forth updates

For further details refer to http://www.w3.org/TR/mmi-arch/.

### E. Mobile Context Capabilities

The restriction on the screen size of mobile phones makes it crucial that the most important and relevant information gets delivered to the customer, or the customer sees those aspects of an app's features that are relevant to his/her and in context. Therefore personal preferences for consumption, personalization of services, personalization based on context (various domains – such as location, social settings, relationships etc.) become center stage.

Key building blocks remain the same the context capabilities can be basic to advanced, and are determined through the elements included, we define these as optional elements that can be selected by the product vendor, implementers based on the capability level desired. The key building blocks are:

- Context Derivation Model: Data Model used as the basis for determining the context of the user.
- Context Rules: Rules (either user specified or derived through various mechanisms) that govern the data model
- Context Inference Engine: Processes the contextual data generated from the model and governed by the rules to infer events in the form of service calls that process the context optimized data.

These essential elements are represented in the figure 4 below.

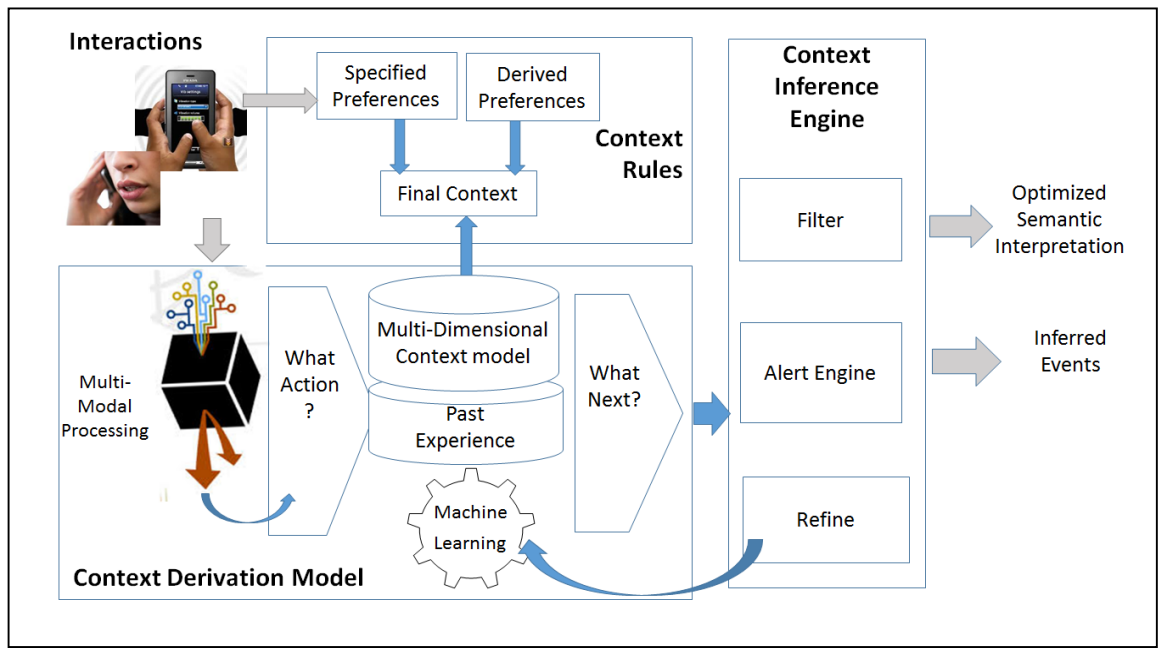

Fig. 4. Building blocks for Mobile Interaction Capability

The capability of the context model may vary from simple User Profile based context derivation to Application based context derivation or may contain multiple parameters that go together to determine the Context Model.

- **Personalization driven by User Profile:** Targeted, contextual personalization based on user profile data are common place. Profile data includes user data gathered across various channels and centralized into the user profile. In the enterprise informational reporting has become the norm for sales/field executives with limited access to the company intranet personalized information alerts either from the backend ERP or KPI's based alerts from BI systems, or personalized alerts from CRM are common place, and dramatically increase office productivity. Increasingly, top management in an organization relies on KPI or event driven personalized information updates to take key decisions. Other common applications in the enterprise also include workflow approvals through the mobile, where workflows are driven through the User Profile available to the ERP application. Creating a customized app experience depends on audience segmentation and the ability to build user profiles and the ability to analyze demographic characteristics based on this.

    Research on content adaption framework for browser based applications has been carried out and is referenced in [3]. Standards describing mobile delivery context have evolved Composite Capabilities/Preference Profiles (CC/PP) by W3C and the User Agent Profile (UAProf) by the Open Mobile Alliance. UAProf had to define functionality that CC/PP left unspecified and has been more successful and therefore adoptions have been incorporated by some vendors, these are one-off adoptions and not the industry norm. UAProf functionality makes it possible for servers to get information about the phone capabilities that can be hardware such as screen size, or software example name and version of the Web browser and details about the various installs. Agent string on the handset sends the information to server, which is able to identify the profile file and hence the capability information for the handset. References [16], [17] and [18] cover the research work carried out.

- **Personalization driven by Application Usage:** Application behavioral data represents how user has interacted with the application over time. Expanding into analysis of application usage and consumer behaviors across multiple channels helps deliver detailed information and makes it possible to reduce application clutter and increase the user experience.

- **Multi-Dimensional Profiling:** To succeed in marketing global brands need to focus on a number of aspects to such as globalization across geographies, blended with localization along with personalization. Usage of Multi-Dimensional Context Model is recommended as it ensures flexibility for future accommodation of additional contextual parameters. Hence, we recommend this as an element of the reference framework.

    Extensive work has been carried out on mobile personalization and we refer to some of these in reference [14] and [15].

- **Multimodal Processng:** Multi-modal interaction including visual, sound and speech related interactions. Through these modalities, it is possible to involve all the five senses (seeing, hearing, smelling, tasting and feeling) in the interaction. When more than one modal interaction is involved they need to be processed together to determine the users actual response in its entireity.

- **Predective / Past Experience:** In certain cases by analyzing historical data user can be prompted for next bext action.

- **Machine Learning:** Some very simple ones today include personalization through questions on preferences and search based on voice interaction whereas many special purpose applications have emerged that leverage intelligent agents for machine learning to achieve hyper-personalization. Ongoing research in the arena focuses on creation of agents with ability to execute intelligent action or multi-agents that are able to interact with other agents to carry out tasks.

Mobile personalization includes the ability for carrying out efficient filtering for showing relevant information thereby leading to intelligent contextual and decreased data transmission, that learns from prior experience of both user driven application usage and search / information retrieval. Athoritative research has been carried out in the area we reference this in [19], [20] and [21].

## F. Mobility Services Capabilities

Mobile application development is fundamentally different from traditional desktop or web applications, mobile applications are designed to exploit the capabilities of mobile devices (such as GPS to determine geopositioning and near field communication).

Enterprise mobile backend as a service (mBaaS) completely abstracts the server-side infrastructure. An MBaaS platform typically offers the following features/services:

- Push notifications
- Online and offline workflows
- DataStore API & binary storage
- Integration with social networking sites
- Secure connectivity
- Device syncing and caching
- Ability to automatically generate interfaces (REST based)

mBaaS is a suitable building block that has reached a level of maturity therefore it can serve as a point of reference,. Besides this other building blocks such as Services collaboration, User management, Data management are also essential aspects that provide support capabilities to enable mobility services.

## G. Capabilities for Quality-Of-Service – Computation & Connectivity

Current mobile devices are constrained in their capability to deliver performance and using multiple apps often leads to user frustration due to hung mobile or unable to load app etc.

Critical applications such as mHealth in the personal space or work space applications involving shop floor/production need to overcome the constraints of the mobile device and network to achieve high computation and/or seamless connectivity, for such applications quality-of-service takes center stage. Overcoming computation and/or connectivity constraints is a basic requirement; personalization is the next level of productivity improvement. Virtualization is a key area where innovation is happening. The concept of hardware resource virtualization in the data center is being extended to network resources. We summarize below a number of innovations that are happening to augment the QoS capability of the mobile device.

- **Application Delivery Controller (ADC):** An ADC is typically placed in a data center between the firewall and one or more application servers in the DMZ. It was primarily meant to provide the capabilities of a load balancer – that deal with front end optimizations rather than backend optimizations. However, many vendors improvised and added additional features such as compression, cache (static data as well as TCP buffering for slow clients), content switching, load balancing connection multiplexing, traffic control, application level security, content manipulation and context aware switching, denial-of-service protection, advanced routing strategies, monitoring and load management and the feature of priority queuing besides the basic functions of TCP Buffering, TCP offload, HTTP Security etc. There are two optimization paths that ADC innovations have adopted general network optimization and application/framework specific optimizations (optimization strategies that work best with a particular application framework).

- **Mix of Physical ADC and Virtual ADC (vADC):** Virtual ADC is a combination of physical ADC customer specific vADC deployed in the application delivery tier. Scenarios which are highly dynamic in nature and require flexibility and rapid scalability for specific application workloads where the application requires complex and compute intensive workloads.

- **Mobile Cloud Computing (MCC):** Cloud computing provides on demand elastically of infrastructure, platforms and software on the cloud. 200+ million businesses have plans to leverage mobile cloud services by this year alone; this is expected to push the revenue in this area to over $5 billion. Some advantages that MCC brings include scalability, shared resources and dynamic on-demand provisioning of shared resources on a self-service basis along with ease of integration with 3$^{rd}$ party providers. M-commerce is a key area where innovations are already underway and we expect the impact to be seen in mobile financials, advertising, shopping etc. A number of authoritative papers on MCC are available in Reference [1], [4], [6] and [8].

- **CloneCloud and Cloudlets:** Are further improvisations on MCC that reduce the network delay by using the nearby computers or data centers. They enable cloning of data and applications from the mobile to the cloud and selectively enable execution of operations delivering back results back on the mobile.

- **Predictive Provisioning:** One-of innovations on top of cloud-optimized ADC are happening that enable further optimizations. Such as that by Brocade, that has lately introduced mechanism that assesses the connections per second and CPU impact on servers before application gets launched.

- **Optimized Hosting:** Similar to the techniques being used for hyper-personalization multi-dimensional context model and machine learning are being researched extensively to determine possible dynamic cloud optimized hosting methods.

- **Network Function Virtualization (NFV):** In late 2012 Industry specification Group (ISG) within the European Telecommunications Standards Institute (ETSI) conceptualized Network Functions Virtualization (NFV) requirements (not standards, efforts are underway to establish an NFV ecosystem). NFV assists in realizing greater benefit in enhancing service delivery and reducing overall costs through managing closed and proprietary appliances on the network through network virtualization with the intention of optimizing the deployment of network functions (such as firewalls, DNS, load balancers etc.). Virtualization of network resource without worrying about where it is physically located, how much it is, how it is organized, etc. Programing ability to change behavior on the fly, dynamic Scaling, automation, visibility to resources and connectivity, ability to optimize network device utilization are some of the key benefits that NFV is expected to bring.

- **Software Defined Networking:** SDN goes beyond what NFV does by optimizing the deployment of the underlying networks. NFV and SDN are complementary. It is based on the concept of logically centralized intelligence facilitated by a global or domain view of the network vs. autonomous systems where nodes are unaware of the overall state of the network. Software defined network is a new architecture that enables more agile and cost effective networks and enables Network Function Virtualization towards realizing greater benefit in enhancing service delivery and reducing overall costs. Open Network Foundation (ONF) taking a lead in standardization of SDN. Both SDN and NFV will go hand in hand to ensure agile and cost effective networks.

- **Future Focus 5G:** Constraints associated with the current 4G Networks limits the capability of the mobile, research on the 5G network is underway and a framework is evolving which is expected to address the elements of :
    - Network interoperability
    - Increased computation through dynamic selection of channels
    - Network predictions
    - Seamless connectivity
    - Transparent selection of best available choice
    - Dynamic spectrum allocation

Additionally the 5G network is also expected to provision for geo-location databases, optimization parameters, network detection and distributed resource management. In the interim that 5G becomes a reality various vendor initiatives and academia lead research are driving innovations that may selectively be incorporated into the various layers of the 5G framework. Number of authoritative papers exploring various areas and frameworks fewer than 5G are referenced in [11], [12] and [13].

### H. Capabilities desired for Interoperability

Interoperation on the mobile platform has two areas where capability is desired:

- Frontend where any App to any device capability should exist so that an application once developed can be run on multiple platforms, this is particularly important for a services organization in the business of writing apps for multiple devices. There are three types of apps that can be developed Native Apps (specific to the mobile platform), HTML5 apps that use standard web technologies and therefore work on multiple devices and hybrid apps (embed HTML5). Though native apps are the fastest and give the best performance we recommend Hybrid Apps that will allow apps to be ported across devices. Reference [24] covers an extensive discussion on the issue of interoperability for the mobile.

- App as a Service – In addition it should also be possible for apps to be delivered as services, where multiple services can be composed together to achieve outcomes. Adherence to "Mobile backend as a Service" (mBaaS) will enable app to app communication. Therefore we recommend this as a key building block to enable the interoperation capability.

I. **Capabilities desired for Security**

The requirement of highly mobile workers for hyper-personalization and hyper quality-of-service is not without increased security risks. Ability to manage security risks for enterprise mobile enablement can enable IT to meet the requirements of both telework and BYOD in enterprises. Recent incidents where well known Hollywood personalities have had mobile phone hacked and personal data stolen have shown us the need to manage security at all levels – from the physical device, to the data under transmission as well as data stored, additionally if cloud provisioning is involved then multi-tenancy security issues etc.

As mobile based services increase, security threats will also increase. The network is dispersed over large areas, exposing them to possibility of hackers who can reprogram various aspects of the network. Networks need to ensure resilient operation despite the presence of a small number of Trojan network nodes. It is extremely difficult to defend against denial-of-service attacks. Denial-of-service attacks can occur at any layer; therefore decentralized mechanisms for security are being extensively researched.

The extensive researched and solutions carried out are referenced in authoritative papers on the topic under [2], [5], [7], [9], [10], [23], [25] and [26].

## 3.CONCLUSION

We delve on the phenomenal progress made in the area of mobile computing in this decade and the multiple innovations that are quickly emerging. In the coming next few years mobile computing is expected to emerge with key dimensions elaborated in the paper expected to take center stage, the movement will be driven more by the consumer needs and vendors and standards body need to quickly catch up to ensure interoperability does not get left behind. We define a framework for mobile as a secure personal assistant that brings together the various building blocks. The reference architecture defined above serves the following key principles:

- It is a generic vendor neutral framework
- It is either mapped to existing mature standards of compliance or helps identify areas where standards exist
- It is capable of being instantiated to produce both industry architectures and solution architectures